\documentclass{article}

\usepackage[utf8]{inputenc}
\usepackage{xcolor}
\usepackage{amsmath,amssymb}
\usepackage{mathtools}
\usepackage{mwe}
\usepackage{subfig}
\usepackage{float}
\usepackage{hyperref}
\hypersetup{
	colorlinks = true,
	linkcolor = red,
	anchorcolor = blue,
	citecolor = blue,
	filecolor = blue,
	urlcolor = blue}
\usepackage[top=2cm,bottom=2cm,right=2cm,left=2cm]{geometry}

\title{An operator approach to the Madelung–Bohm continuity equation}

\author{M. A. García-Márquez$^{1}$, H. M. Moya-Cessa$^{1}$, I. Ramos-Prieto$^{1 }$, F. Soto-Eguibar $^{1}$\\
$^{1 }$ Instituto Nacional de Astrof\'isica, \'Optica y Electr\'onica, Tonantzintla, Puebla.}

\date{\today}

\begin{document}

\maketitle

\section*{Abstract}
We solve the probability continuity equation within the Madelung–Bohm framework, assuming a separable phase expressed as $S(x,t) = Q(x)\dot{\nu}(t) + \mu(t)$. Using operator methods, we reformulate the wave function's amplitude into a form that is more practical for application to any given initial condition. This results in the function $F(x,t)$, which characterizes the wavefunction's amplitude, with $F$ being the transformation of the position operator via a squeeze-like operator. To demonstrate how this result can be applied, we examine two scenarios: one where the external potential is zero, causing the dynamics to originate only from the Bohm potential, and another where the form of $F$ mirrors the characteristics of wave propagation within optical waveguide arrays. In the second scenario, it is notable that the amplitude, phase, and potentials might become complex yet still comply with the Schrödinger equation.

\section{Introduction}
In the Madelung–Bohm formulation of quantum mechanics \cite{Madelung (1927), Bohm (1952),Holland (1995)}, originally introduced by Madelung in the context of his hydrodynamic interpretation \cite{Madelung (1927)} and later rediscovered by David Bohm within his alternative interpretation of quantum theory based on hidden variables \cite{Bohm (1952)}, the wave function is expressed in polar form as $\Psi(x,t) = A(x,t)\exp[iS(x,t)]$, with $A$ and $S$ being real-valued functions. Substituting this expression into the Schrödinger equation yields a system of two coupled differential equations: a continuity equation, which describes probability conservation, and a Hamilton-Jacobi type equation, which includes, in addition to the classical potential $V$, an extra term known as the Bohm quantum potential $V_B$. This formulation, through the Hamilton–Jacobi equation, defines an ensemble of possible trajectories for a particle associated with the system, allowing the exploration of various physical scenarios by appropriately choosing the potential. For example, one can reproduce the dynamics equivalent to those of a free particle \cite{Hojman and Asenjo (2020b)}, or even construct situations in which classical and quantum trajectories coincide \cite{Makowski and Konkel (1998)}. More recently, this formalism has also been extended to the context of paraxial optics \cite{Silva-Ortigoza et al. (2022), Silva-Ortigoza et al. (2023)}.

A convenient way to obtain the amplitude $A$ and phase $S$ of the wave function is by solving the Schrödinger equation and identifying both functions through its polar representation. An alternative approach was proposed in \cite{Hojman and Asenjo (2020), Villaflor et al. (2024)}, where a potential field $f$ is defined, which satisfies an integro-differential equation and determines both the amplitude and the phase of the wave function. Another possibility is to directly solve the coupled differential equations arising from the Madelung-Bohm formalism by assuming an ansatz for the phase of the form $S(x,t) = Q(x)\dot{\nu}(t) + \mu(t)$, from which the amplitude can then be determined. This strategy has been used, for example, to derive the Bohm quantum potential associated with a harmonic oscillator with time-dependent frequency \cite{Soto-Eguibar et al. (2021)}, and to provide a physical explanation for the origin of the Gouy phase in the propagation of paraxial beams, due to the mathematical similarity between the Schrödinger and the paraxial wave equations \cite{Moya-Cessa et al. (2022)}.

In this work, we obtain the wave function by solving the probability continuity equation arising from the Madelung-Bohm formalism, assuming a wave function phase of the form $Q(x)\dot{\nu}(t) + \mu(t)$, where $Q(x)$ is a function of position only, and $\nu(t)$ and $\mu(t)$ are time-dependent functions. Unlike the approaches in \cite{Soto-Eguibar et al. (2021), Moya-Cessa et al. (2022)}, we employ an operator-based method through which the expression obtained for the amplitude of the wave function is rewritten in a factorized form. By strategically inserting identity operators and applying the Hadamard lemma \cite{Louisell (1990)}, we simplify the resulting expression, thus facilitating the application of the operators involved to the initial condition. This method yields an explicit expression for the amplitude of the wave function in terms of a function $F(x,t)$, defined as the transformation of the position operator by a squeezed operator \cite{Loudon and Knight (1987), Yuen (1976), Caves (1981), Moya et al. (2014)}, whose time dependence is governed by the temporal structure of the phase, while its spatial dependence is determined through recurrence relations related to the spatial form of the phase. To demonstrate the flexibility of the solution, we apply it to two cases. The first considers a vanishing external potential and a wavefunction with a quadratic phase, where nontrivial system dynamics emerges solely due to the Bohm potential. The second case focuses on a scenario in which the function $F(x,t)$ is obtained by observing that the recurrence relations that define $F$ exhibit a structure analogous to that describing wave propagation in optical waveguide arrays. As a result, we find that the amplitude, phase, potential, and Bohm potential are not necessarily real-valued functions, contrary to what is usually assumed in the Madelung–Bohm formalism. Consequently, the Hamiltonian of the system is non-Hermitian; nevertheless, the Schrödinger equation remains satisfied.

\section{Madelung–Bohm formalism}
We consider the Schrödinger equation (for simplicity, we set $m = 1 $ and $\hbar = 1$)  
\begin{equation}
i\frac{\partial \Psi(x,t)}{\partial t} = -\frac{1}{2}\frac{\partial^2 \Psi(x,t)}{\partial x^2} + V(x,t) \Psi(x,t),
\label{Schrodinger equation}
\end{equation} 
where $\Psi(x,t)$ is the wave function and $V(x,t)$ represents the external potential. Following the Bohm–Madelung formalism, the wave function is expressed in polar form as \cite{Holland (1995)} 
\begin{equation}
\Psi(x,t) = A(x,t) e^{iS(x,t)},
\label{Polar form of the wave function}
\end{equation}
where $A(x,t)$ and $S(x,t)$ are assumed to be real-valued functions of position and time. In what follows, we use the notation $\dot{f} = \frac{\partial f}{\partial t}$ and $f' = \frac{\partial f}{\partial x}$ to denote the temporal and spatial derivatives, respectively. Substituting equation (\ref{Polar form of the wave function}) into the Schrödinger equation yields the quantum Hamilton-Jacobi equation,
\begin{equation}
\frac{1}{2} S'^2 +V+  V_B + \dot{S} = 0,
\label{Quantum Hamilton–Jacobi equation}
\end{equation}  
together with the continuity equation, which expresses the conservation of probability,
\begin{equation}
    \frac{1}{2} \left( 2A'S' + A S'' \right) + \dot{A} = 0,
    \label{Continuity equation 0}
\end{equation}  
where the Bohm potential is defined as
\begin{equation}
    V_B = -\frac{1}{2} \frac{A''}{A}.
\end{equation}

\section{Solution of the continuity equation via operator methods}
The continuity equation (\ref{Continuity equation 0}) may be written as a Schrödinger-like equation
\begin{equation}\label{Continuity equation 1}
    \frac{\partial A}{\partial t}=-\frac{1}{2}\left(2 S' \partial_x+S''\right) A,
\end{equation}
where we use the operator $\partial_x = \frac{\partial}{\partial x}$, which satisfies the commutation relation $[\partial_x, x] = 1$. We assume that the phase of the wave function takes the form 
\begin{equation}  \label{Phase}
S(x,t) = Q(x)\dot{\nu}(t) + \mu(t),
\end{equation}
where $Q$ is a function of $x$ only, while $\nu$ and $\mu$ are functions of $t$ only. Under this assumption, equation~(\ref{Continuity equation 1}) can be rewritten as
\begin{equation} \label{Continuity equation 3}
\frac{\partial A}{\partial t}=-\dot{\nu}\left(Q' \partial_x+\frac{Q''}{2}\right) A,
\end{equation}
which is readily solvable with solution
\begin{equation} \label{Continuity equation solution 1}
A(x, t)=\exp\left[{-\nu(t)\left(Q'(x) \partial_x+\frac{Q''(x)}{2}\right)}\right] A_0(x),
\end{equation}
where $A_0(x) = A(x,0)$ denotes the initial condition, an arbitrary square-integrable function of position. From this we conclude that $\nu(0) = 0$ must hold.

To find an expression that can be easily applied to the initial condition to obtain the amplitude of the wave function, we propose the following ansatz, which rewrites equation~(\ref{Continuity equation solution 1}) as
\begin{equation} \label{Ansatz}
    A(x, t)=e^{R(x)} e^{-\nu(t) Q'(x) \partial_x} e^{-R(x)} A_0(x),
\end{equation} 
where $R(x)$ is a function to be determined. We make use of the well-known Hadamard lemma \cite{Louisell (1990)}, which states that if $\hat{A}$ and $\hat{B}$ are two noncommuting operators and $\xi$ is a parameter, then 
$e^{\xi\hat{A}} \hat{B} e^{-\xi\hat{A}} = \hat{B} + \xi[\hat{A}, \hat{B}] + \frac{\xi^2}{2!} [\hat{A}, [\hat{A}, \hat{B}]] + \frac{\xi^3}{3!} [\hat{A}, [\hat{A}, [\hat{A}, \hat{B}]]] + \cdots$, 
which allows us to compute
\begin{equation}\label{Hadamard's lemma}
e^{R(x)}  \partial _x e^{-R(x)}=  \partial _x-  R'(x);
\end{equation}
hence, by using the result from equation~(\ref{Hadamard's lemma}), we can rewrite equation~(\ref{Ansatz}) as
\begin{equation} \label{continuity equation solution 3}
A(x, t)=e^{-\nu(t) Q'(x)\left(\partial_x-R'(x)\right)} A_0(x).
\end{equation}
Comparing equations~(\ref{Continuity equation solution 1}) and~(\ref{continuity equation solution 3}), we find that
\begin{equation}
R^{\prime}(x)=-\frac{Q''(x)}{2 Q'(x)}=-\frac{1}{2}\left(\ln Q'(x)\right)',
\end{equation}
whose solution is
\begin{equation} \label{R(x)}
R(x)=-\frac{1}{2} \ln( Q'(x)) +a,
\end{equation}
where $a$ is an integration constant. Thus, the amplitude given in (\ref{Ansatz}) becomes
\begin{equation} \label{continuity equation solution 4}
    A(x, t)=e^{-\frac{1}{2} \ln (Q'(x))} e^{-\nu(t) Q'(x) \partial_x} e^{\frac{1}{2} \ln (Q'(x))} A_0(x). 
\end{equation}
To apply the derivative operator $\partial_x$, we insert the identity operator expressed as  $e^{\nu(t) Q'(x) \partial_x} e^{-\nu(t) Q'(x) \partial_x}$ into equation~(\ref{continuity equation solution 4}) as follows:
\begin{equation}\label{continuity equation solution 5}
\begin{aligned}
    A(x, t)&=e^{-\frac{1}{2} \ln (Q'(x))} e^{-\nu(t) Q'(x) \partial_x} \tilde{A}_0(x) e^{\nu(t) Q'(x) \partial_x} e^{-\nu(t) Q'(x) \partial_x} 1\\
    & =e^{-\frac{1}{2} \ln (Q'(x))} e^{-\nu(t) Q'(x) \partial_x} \tilde{A}_0(x) e^{\nu(t) Q'(x) \partial_x} 1,
\end{aligned}
\end{equation}
where $\tilde{A}_0(x) = e^{\frac{1}{2} \ln(Q'(x))} A_0(x)$ is a function of
$x$, and we have used the fact that $e^{-\nu(t) Q'(x) \partial_x} 1 = 1$. Applying the Hadamard lemma, we find that
\begin{equation}  \label{squeeze-like operator}
    e^{-\nu(t) Q'(x) \partial_x} x e^{\nu(t) Q'(x) \partial_x}=F(x,t),
\end{equation}
where 
\begin{equation} \label{Function F}
    F(x,t)=\sum_{k=0}^\infty\frac{(-\nu(t))^k}{k!}f_k(x),
\end{equation}
and the functions $f_k$ are given by 
\begin{equation} \label{compact form}
     f_0=x; \qquad f_{k+1}=Q'\frac{df_k}{dx}, \quad k=0,1,2,\dots .
\end{equation}
We observe that eqs.~(\ref{squeeze-like operator}),~(\ref{Function F}), and~(\ref{compact form}) define the action of a squeeze-like operator~\cite{Moya et al. (2014)} with a time-varying squeezing parameter on the position operator, and, in particular, if $Q(x) \propto x^2$, we obtain the usual squeezing operator~\cite{Loudon and Knight (1987), Yuen (1976), Caves (1981)}; therefore,
\begin{equation} \label{eq aux}
    e^{-\nu(t) Q'(x) \partial_x} \tilde{A}_0(x) e^{\nu(t) Q'(x) \partial_x}=\tilde{A}_0(F(x,t)).
\end{equation}
Finally, using equation~(\ref{eq aux}) and performing the necessary mathematical simplifications, we find that the amplitude of the wave function (\ref{continuity equation solution 5}) is given by
\begin{equation} \label{continuity equation solution 7}
    A(x, t)=\left(\frac{Q'(F(x,t))}{Q'(x)}\right)^\frac{1}{2}A_0(F(x,t)), 
\end{equation}
where $F(x,t)$ is defined in equation (\ref{Function F}). In this way, choosing the amplitude $A(x,t)$ as in equation~(\ref{continuity equation solution 7}), ensures that the probability continuity equation~(\ref{Continuity equation 0}) is automatically satisfied. However, it is still necessary to appropriately determine or choose the functions $Q(x)$, $\nu(t)$, $\mu(t)$, and $V(x,t)$ so that they satisfy the quantum Hamilton–Jacobi equation~(\ref{Quantum Hamilton–Jacobi equation}). This freedom allows us to study various cases of the problem, some of which are presented in the following section.

\section{Effects of the Bohm quantum potential for a free particle}
The aim is to study the case in which the potential is zero, $V(x,t) = 0$, while the Bohm potential remains nonzero, $V_B(x,t) \neq 0$. We consider the phase $S(x,t)$ to be quadratic in position, that is, $Q(x) = \kappa \frac{x^2}{2}$, where $\kappa$ is a real constant. This choice leads to $F(x,t) = x e^{-\kappa\nu(t)}$, as expected.

Based on this, we can determine the amplitude of the wave function as
\begin{equation}
    A(x,t) = e^{-\frac{\kappa\nu(t)}{2}} A_0(x e^{-\kappa\nu(t)}),
\end{equation}
 accordingly, the Bohm potential is found to be
\begin{equation}
    V_B(x,t) = -\frac{e^{-2\kappa\nu(t)}}{2 A_0(x e^{-\kappa\nu(t)})} \frac{d^2 A_0(x e^{-\kappa\nu(t)})}{d(x e^{-\kappa\nu(t)})^2}.
\end{equation}

To explicitly obtain the amplitude and associated Bohm potential, we consider the initial condition $A_0(x) = e^{-\eta x^2}$, with $\eta$ a real constant, so that the function is square-integrable. Under this assumption, the Bohm potential takes the form
\begin{equation}
    V_B(x,t) =\eta \left(1 - 2\eta x^2 e^{-2\kappa\nu(t)}\right) e^{-2\kappa\nu(t)}.
\end{equation}

Next, we determine the functional form of $\nu(t)$ and $\mu(t)$ by requiring the quantum Hamilton-Jacobi equation to be satisfied in the case where the potential is zero, that is, $V(x,t) = 0$. This leads to the following expression:
\begin{equation}
    \frac{\kappa x^2}{2} \left(\kappa  \dot{\nu}^2 + \ddot{\nu} - \frac{4\eta^2}{\kappa} e^{-4\kappa\nu} \right) + \dot{\mu} + \eta e^{-2\kappa\nu} = 0.
\end{equation}

Due to the independence between position $x$ and time $t$, it follows that the above equation is satisfied if and only if the following conditions hold:
\begin{equation}
    \kappa  \dot{\nu}^2 + \ddot{\nu} - \frac{4\eta^2}{\kappa} e^{-4\kappa\nu} = 0,
    \label{Differential equation nu}
\end{equation}
\begin{equation}
    \dot{\mu} + \eta e^{-2\kappa\nu}= 0.
    \label{Differential equation mu}
\end{equation}
The solution to the differential equation~(\ref{Differential equation nu}) is
\begin{equation}
    \nu(t) = \frac{1}{2\kappa } \ln{\left[ \frac{4\eta^2 + \kappa^4c_1^2 (t - c_2)^2}{\kappa^2c_1} \right]},
\end{equation}
and for the differential equation~(\ref{Differential equation mu}) we obtain
\begin{equation}
    \mu(t) = -\frac{1}{2} \arctan\left[ \frac{\kappa^2 c_1}{2\eta} (t - c_2) \right] + c_3,
\end{equation}
where $c_2 = \pm\frac{\sqrt{\kappa^2 c_1 - 4\eta^2}}{\kappa^2 c_1}$, in order to meet the requirement that $\nu(0) = 0$; the constants $c_1$ and $c_3$ are fixed by the initial conditions of the wave function.

Finally, the amplitude, phase, and Bohm potential are given by
\begin{equation}
    A(x,t) = e^{-\frac{\kappa\nu}{2}-\eta x^2e^{-2\kappa \nu}} =\left( \frac{\kappa ^2c_1} {4\eta^2 + \kappa^4c_1^2 (t-c_2)^2}\right)^{1/4}e^{-\frac{\kappa^2c_1 \eta x^2}{4\eta^2 + \kappa^4c_1^2 (t-c_2)^2}},
\end{equation}
\begin{equation}
    S(x,t) = \frac{\kappa^4c_1^2 ( t-c_2 ) x^2}{2(4\eta^2 + \kappa^4c_1^2 (t-c_2 )^2)} - \frac{1}{2}\arctan\left( \frac{ \kappa^2c_1 }{2\eta}(t-c_2 ) \right) +c_3,
\end{equation}
\begin{equation}
    V_B(x,t) = \frac{\eta \kappa ^2 c_1 \left( 4\eta^2 +\kappa^4 c_1^2 (t-c_2)^2 - 2\eta\kappa^2 c_1 x^2 \right)}{\left( 4\eta^2 + \kappa^4c_1^2 (t-c_2 )^2 \right)^2}.
\end{equation}
The Bohm potential then represents a repulsive time-dependent harmonic oscillator, the expressions for the amplitude and the phase agree with those of free propagation of a Gaussian profile, and the $\arctan$ in $S(x,t)$ is nothing but the Gouy phase \cite{Moya-Cessa et al. (2022)}. From these expressions, it is evident that the constant $c_1$ is a parameter that controls the spatial width of the Gaussian function that describes the amplitude of the wave function. The constant $c_2$ determines the time at which the Gaussian width reaches its minimum, and finally, $c_3$ represents a constant phase. Figures~\ref{fig:(eta,k,c1,c2,c3)=(0.1,0.5,0.8,2.,0)} and~\ref{fig:(eta,k,c1,c2,c3)=(0.1,0.5,0.2,2.,0)} show the behavior of the probability density, phase, and Bohm potential for the parameter sets $(\eta, \kappa, c_1,c_2,c_3) = (0.1,0.5,0.8,2,0)$ and $(\eta, \kappa, c_1,c_2,c_3) = (0.1,0.5,0.2,2,0)$, respectively.

\begin{figure}[H]
    \centering
    \includegraphics[width=0.95\linewidth]{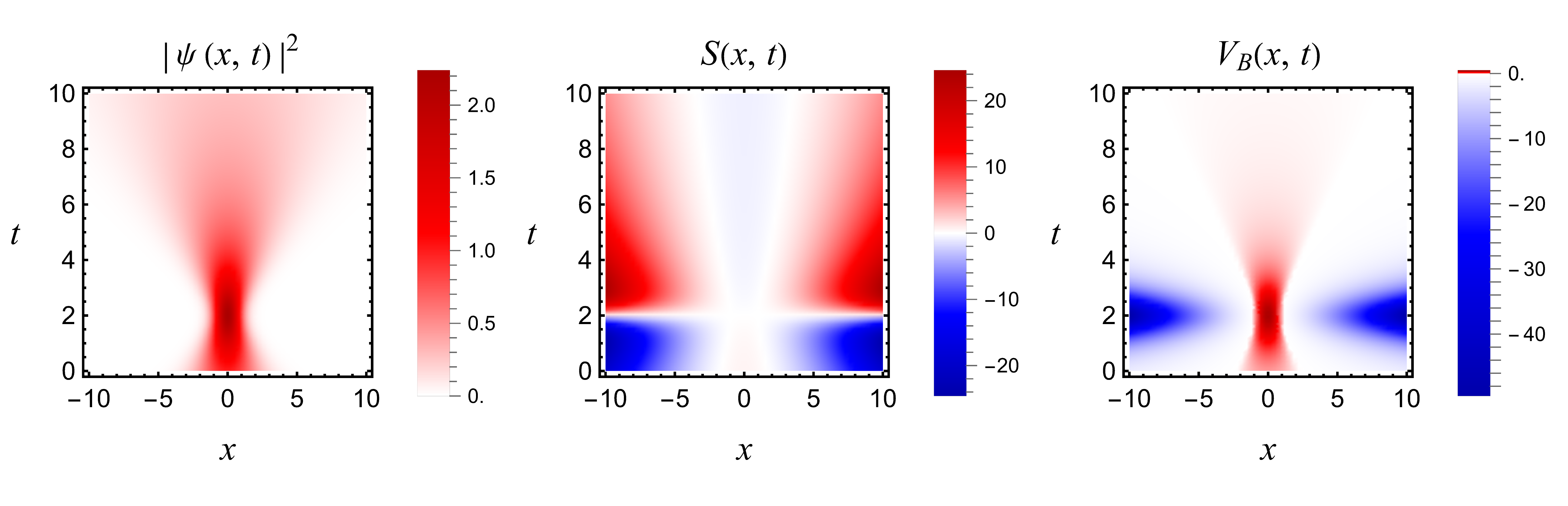}
    \caption{Density plot of the wave function for the case $(\eta, \kappa, c_1,c_2,c_3) = (0.1,0.5,0.8,2,0)$. On the left, the probability density $|\psi(x,t)|^2$ is shown; in the middle, the phase $S(x,t)$; and on the right, the Bohm potential $V_B(x,t)$.}
    \label{fig:(eta,k,c1,c2,c3)=(0.1,0.5,0.8,2.,0)}
\end{figure}
\begin{figure}[H]
    \centering
    \includegraphics[width=0.95\linewidth]{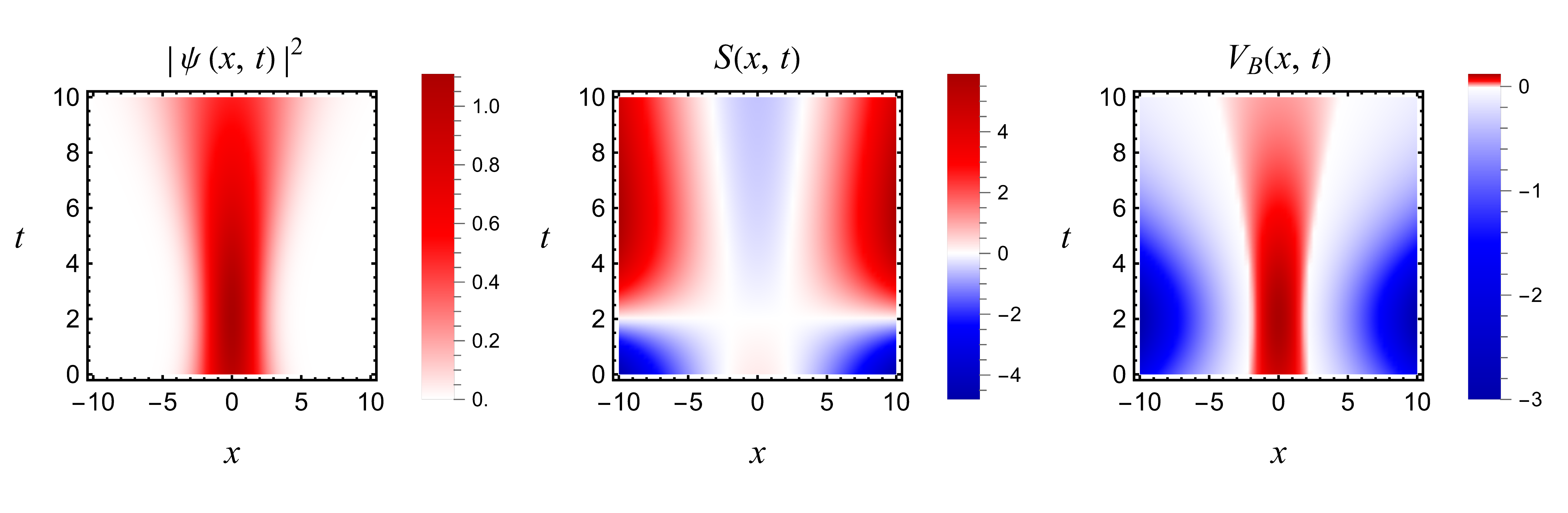}
    \caption{Density plot of the wave function for the case $(\eta, \kappa, c_1,c_2,c_3) = (0.1,0.5,0.2,2,0)$. On the left, the probability density $|\psi(x,t)|^2$ is shown; in the middle, the phase $S(x,t)$; and on the right, the Bohm potential $V_B(x,t)$.}
    \label{fig:(eta,k,c1,c2,c3)=(0.1,0.5,0.2,2.,0)}
\end{figure}

It is particularly interesting to observe that the probability density evolves over time, despite the absence of any external potential $V(x,t)$ acting on the system. This indicates that the changes in the wave function's behavior are exclusively driven by the Bohm potential.

\section{Waveguide-array-like equations}
Equation \eqref{compact form} presents recurrence relations that exhibit a structure similar to that which describes field propagation within a waveguide array \cite{Tapia-Valerdi et al. (2025a),Tapia-Valerdi et al. (2025), Perez-Leija et al. (2010)}. In the following, we provide an example of the behavior of the wave function and, consequently, of the amplitude~(\ref{continuity equation solution 7}) and the phase~(\ref{Phase}), in the case where the function $F(x,t)$ is obtained by modeling the equations in~(\ref{compact form}) analogously to the propagation of light in waveguide arrays; that is, through the differential equation
\begin{equation}
    \frac{d|f\rangle}{dx}=\frac{\hat{V}}{Q'} |f\rangle,
    \label{Differential equation for f's}
\end{equation}
where
\begin{equation}
    |f\rangle = \sum_{k=0}^\infty f_k(x)|k\rangle =\begin{bmatrix}
f_0 \\
f_1 \\
\vdots \\
\end{bmatrix}, \qquad  \qquad 
\hat{V} =\sum_{k=0}^\infty|k\rangle\langle k+1|=
\begin{pmatrix}
0 & 1 & 0 & 0 & \cdots \\
0 & 0 & 1 & 0 & \cdots \\
0 & 0 & 0 & 1 & \cdots \\
\vdots & \vdots & \vdots & \vdots & \ddots
\end{pmatrix},
\end{equation}
where $\{ |k\rangle \}$ denotes the number basis,  with $|k\rangle$ representing a column vector whose components are zero everywhere except at the $k$th position, and $\hat{V}$ is the well-known London-Susskind–Glogower operator \cite{Susskind and Glogower (1964), London (1926)}; in our case, the condition $f_0 = x$ must be satisfied. The solution to the differential equation~(\ref{Differential equation for f's}) is
\begin{equation}
    |f\rangle=e^{\hat{V}\int \frac{dx}{Q'}}|f(0)\rangle.
\end{equation}
To proceed, we assume that $|f(0)\rangle = |n\rangle$, which satisfies $\hat{V} |n\rangle = |n-1\rangle$. It then follows that
\begin{equation}
 |f\rangle=\sum_{k=0}^n\frac{\left(\int \frac{dx}{Q'}\right)^{n-k}}{(n-k)!}|k\rangle,   
\end{equation}
or, equivalently,
\begin{equation}
f_{k} = \frac{1}{(n-k)!} \left(\int\frac{dx}{Q'}\right)^{n-k} 
\label{f's 1},
\end{equation}
with $k = 0, 1, \ldots, n$. In order to determine the functional form of $Q$, we focus on the case $k = 0$ and use condition $f_0 = x$ to obtain
\begin{equation}
f_{0} = \frac{1}{ n !} \left(\int\frac{dx}{Q'}\right)^{n } =x,
\end{equation}
which leads to the solution
\begin{equation}
Q' = \frac{n}{\left(n!\right)^{\frac{1}{n}}} x^{\frac{n-1}{n}},
\label{Q'}
\end{equation}
and, substituting this expression for $Q'$ into equation~(\ref{f's 1}), we find  
\begin{equation}
f_k = \frac{\left(n!x\right)^{\frac{n-k}{n}}}{(n-k)!}.
\end{equation}
Therefore, the explicit expression for the function $F$ can be written as
\begin{equation}
F(x,t) =x \sum _{k=0} ^{n}\frac{n!}{k!(n-k)!}  (-\nu(t)\left(n!x\right)^{-\frac{1}{n}})^k = x\left(1-\frac{\nu(t)}{(n!x)^{1/n}}\right)^n.
\end{equation}
Consequently, given $\nu(t)$ and $\mu(t)$, the amplitude $A$ takes the form of
\begin{equation}
A(x,t) = \left[\frac{F(x,t)}{x}\right]^{\frac{n-1}{2n}} A_0 (F(x,t)),
\end{equation}
while the phase of the wave function is
\begin{equation}
    S(x,t) = \left[\frac{n^2}{(2n-1)\left(n!\right)^{\frac{1}{n}}} x^{\frac{2n-1}{n}}+c\right] \dot{{\nu}}(t) + \mu(t),
\end{equation}
where $c$ is an integration constant resulting from the integration of equation~(\ref{Q'}). To ensure that the Hamilton–Jacobi equation~(\ref{Quantum Hamilton–Jacobi equation}) is satisfied, we choose a potential of the form
\begin{equation}
    V= -\frac{1}{2} S'^2  -V_B - \dot{S}.
\end{equation} 
It is important to note that, due to the functional form of $F(x,t)$, for $n > 1$, this function can take complex values when evaluated at negative values of $x$. Consequently, the amplitude we have obtained is not purely real, as is typically assumed when working within the Bohm formalism, but may become complex for negative values of $x$. Since the chosen potential is expressed in terms of the amplitude $A$, it also becomes complex, which implies that the Hamiltonian of the system is non-Hermitian. Nevertheless, the Schrödinger equation continues to be satisfied. On the other hand, the functional form of the amplitude $A$ is not square integrable due to the factor $F/x$, which diverges at $x = 0$.

\subsection{Case $A_0(x) = \exp(-x^2)$}
We consider the case in which the initial amplitude is given by a Gaussian function of position, that is, $A_0(x) = \exp(-x^2)$. Due to the complicated nature of the expression for the amplitude $A$, and hence for the potential $V$, a numerical approach is adopted.
In general, we assume that $\nu(t) = \sin(t)$ and $\mu(t) = t$.
In the case where $n = 1$, it is found that
\begin{align}
|\psi(x,t)|^2 &= \exp[-2 (x - \sin(t))^2], 
  &\quad S(x,t) &= t + (x+c) \cos(t), \notag \\
V(x,t) &= -\tfrac{5}{4} + 2 x^2 - \tfrac{5}{4} \cos(2 t)+(c - 3 x) \sin(t), 
  &\quad V_B(x,t) &= \cos(2 t) - 2 x^2 + 4 x \sin(t). \label{Equations, case n=1}
\end{align}
It can be observed that all these functions are real throughout space and time, and with this knowledge we obtain the result that both potentials correspond to harmonic oscillators displaced by time-dependent parameters. To visualize the behavior of the solution, Figure~\ref{fig:A0,n,nu,mu=ex2,1,sent,t} displays the probability density $|\psi(x,t)|^2$, the phase $S(x,t)$, the potential $V(x,t)$ and the Bohm potential $V_B(x,t)$, where the observed oscillatory behavior in the probability density arises because a harmonic time-dependent phase was chosen.

\begin{figure}[H]
\centering
\includegraphics[width=0.63\linewidth]{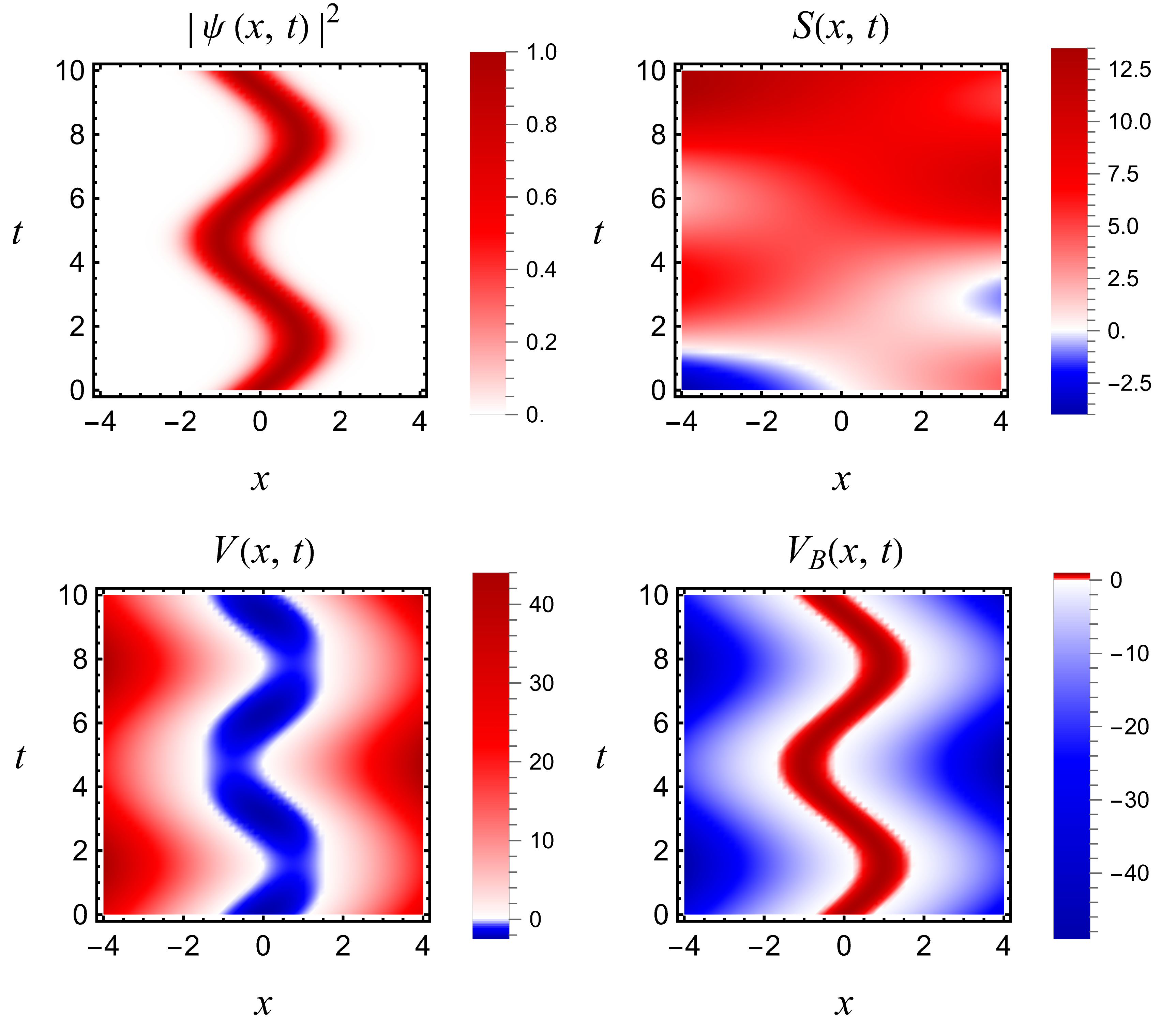}
\caption{Density plots of the probability density $|\psi(x,t)|^2$, the phase $S(x,t)$, the potential $V(x,t)$, and the Bohm potential $V_B(x,t)$ for the case $n = 1$ and $c=0$.}
\label{fig:A0,n,nu,mu=ex2,1,sent,t}
\end{figure}

On the other hand, a more interesting situation arises when the case $n = 2$ is considered. In Figures~\ref{fig:A0,n,nu,mu=ex2,2,sent,t} and \ref{fig:A0,n,nu,mu,x=ex2,2,sent,t,-}, the functions that describe the wave function are shown at intervals $(0.1, 4]$ and $[-4, -0.1)$, respectively, within the time interval $(0, 10)$. 
Note that in Figure~\ref{fig:A0,n,nu,mu=ex2,2,sent,t}, the plots of $\operatorname{Re}\lbrace V(x,t)\rbrace$ and $\operatorname{Re}\lbrace V_B(x,t)\rbrace$ contain black regions where the functions were not plotted. This omission was made because, in those regions, the functions diverge and yield very large values, which would diminish the contrast in the rest of the domain.

Note that in this case, although we have complex functions for $A(x,t)$ and $S(x,t)$ they still obey the equations (\ref{Quantum Hamilton–Jacobi equation}) and (\ref{Continuity equation 0}), but they are outside of the Madelung-Bohm formalism in the sense that they are not real any more for $n>1$ and we would be in the realm of non-Hermitian quantum mechanics. In fact, as shown in the figures \ref{fig:A0,n,nu,mu=ex2,2,sent,t} and \ref{fig:A0,n,nu,mu,x=ex2,2,sent,t,-}, while the functions are purely real in the positive position region, in the negative domain the amplitude, phase, potential, and Bohm potential acquire complex values that lead to probabilities that may exceed one. The functions are not evaluated at $x=0$, where the solution diverges. Therefore, we have found a set of solutions to the Schrödinger equation for non-Hermitian Hamiltonians, with larger values of $n$ yielding longer analytical expressions and more elaborate dynamical behavior.

\begin{figure}[H]
    \centering
    \includegraphics[width=0.95\linewidth]{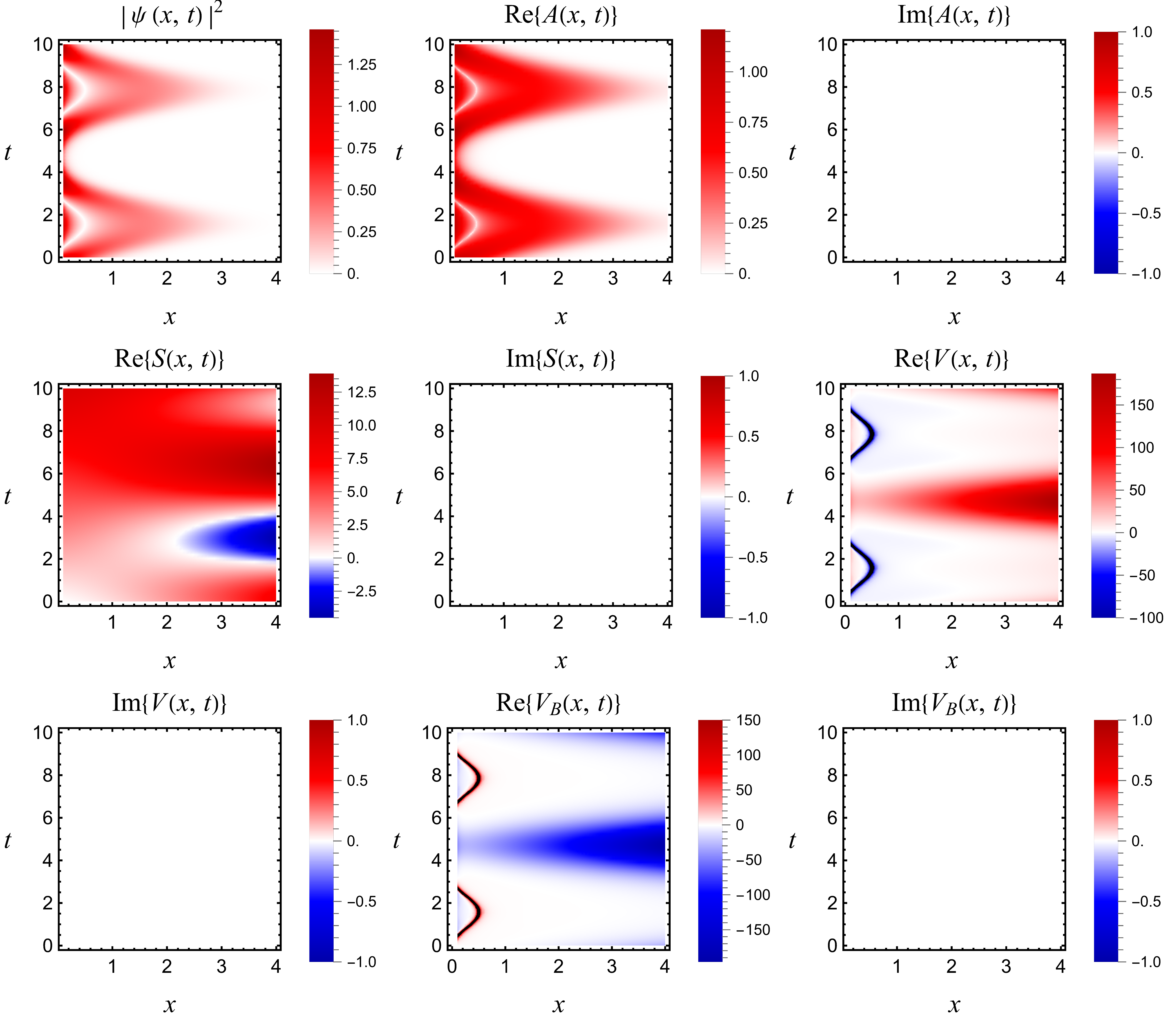}
    \caption{Density plots of the probability density $|\psi(x,t)|^2$, the real and imaginary parts of the amplitude $A(x,t)$, the phase $S(x,t)$, the potential $V(x,t)$, and the Bohm potential $V_B(x,t)$ for the case $n = 2$ in the positive position interval.}
    \label{fig:A0,n,nu,mu=ex2,2,sent,t}
\end{figure}

\begin{figure}[H]
    \centering
    \includegraphics[width=0.95\linewidth]{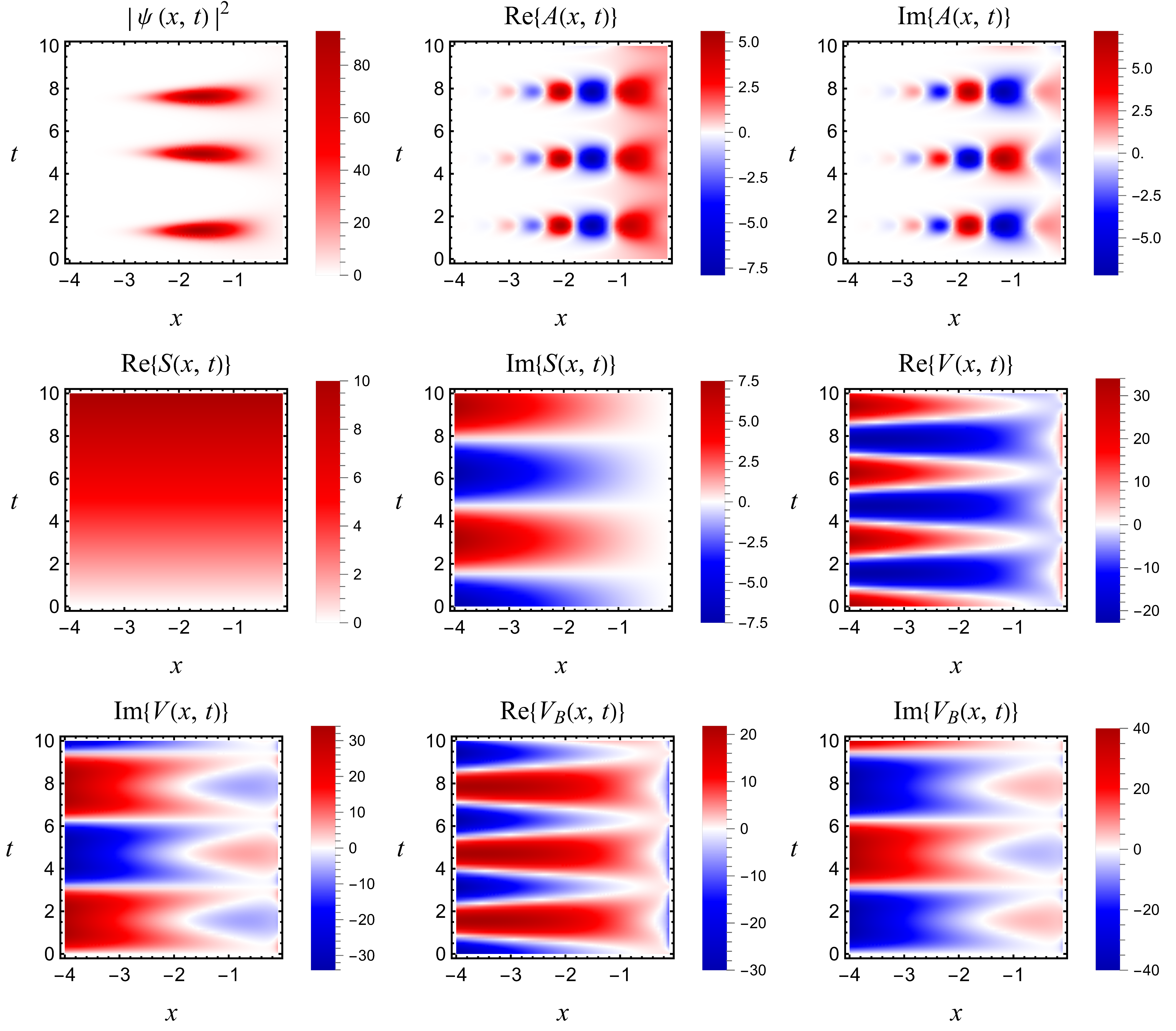}
    \caption{Density plots of the probability density $|\psi(x,t)|^2$, the real and imaginary parts of the amplitude $A(x,t)$, the phase $S(x,t)$, the potential $V(x,t)$, and the Bohm potential $V_B(x,t)$ for the case $n = 2$ in the negative position interval.}
    \label{fig:A0,n,nu,mu,x=ex2,2,sent,t,-}
\end{figure}

\section{Conclusions}
In this work, we have developed an operator-based method to solve the probability continuity equation within the Madelung-Bohm framework, under the assumption of a separable phase of the form $S(x,t) = Q(x)\dot{\nu}(t) + \mu(t)$. This approach not only yields an explicit expression for the wave function amplitude in terms of a function $F(x,t)$, defined through the action of a squeeze-like operator on the position operator, but also reveals the underlying algebraic structure governing the system's dynamics. The operational factorization and use of the Hadamard lemma facilitate the application of the formalism to arbitrary initial conditions, generalizing and unifying previous treatments.

Two representative cases were examined to illustrate the versatility of the method. In the first, in the absence of an external potential ($V = 0$), the time evolution is entirely governed by the Bohm quantum potential, demonstrating how the latter can induce nontrivial dynamics; such as the free propagation of a Gaussian packet with a Gouy phase; even in seemingly simple systems. In the second case, by modeling the recurrence relations for $F(x,t)$ in analogy with those describing optical waveguide arrays, we found that the amplitude, phase, and associated potentials can become complex, leading to non-Hermitian Hamiltonians that nevertheless satisfy the Schrödinger equation. This result significantly extends the scope of the Madelung-Bohm formalism, traditionally restricted to real-valued functions, and establishes a bridge to non-Hermitian quantum mechanics.

These findings underscore the utility of the proposed operator-based method for exploring quantum systems with equivalent hydrodynamic dynamics, both in Hermitian and non-Hermitian contexts. The explicit connection with squeezing operators and the analogy with photonic systems open new avenues for applications in quantum optics, quantum information, and the study of open systems, where non-Hermiticity and complex phases play fundamental roles. 





\begin{thebibliography}{99}

\bibitem{Madelung (1927)} Madelung, E. (1927). Quantum theory in hydrodynamical form. z. Phys, 40, 322.

\bibitem{Bohm (1952)} Bohm, D. (1952). A suggested interpretation of the quantum theory in terms of ``hidden" variables. I. Physical review, 85(2), 166.

\bibitem{Holland (1995)} Holland, P. R. (1995). The quantum theory of motion: an account of the de Broglie-Bohm causal interpretation of quantum mechanics. Cambridge university press.
\bibitem{Hojman and Asenjo (2020b)} Hojman, S. A., \& Asenjo, F. A. (2020). Quantum particles that behave as free classical particles. Physical Review A, 102(5), 052211.
\bibitem{Makowski and Konkel (1998)} Makowski, A. J., \& Konkel, S. (1998). Identical motion in classical and quantum mechanics. Physical Review A, 58(6), 4975.

\bibitem{Silva-Ortigoza et al. (2022)} Silva-Ortigoza, G., Ortiz-Flores, J., Sosa-Sánchez, C. T., \& Silva-Ortigoza, R. (2022). Mechanical properties of the particle associated with the Laguerre–Gauss beams via the quantum potential point of view. Journal of the Optical Society of America B, 40(1), 215-223.

\bibitem{Silva-Ortigoza et al. (2023)} Silva-Ortigoza, G., Julián-Macías, I., Espíndola-Ramos, E., \& Silva-Ortigoza, R. (2023). Exact and geometrical optics energy trajectories in Bessel beams via the quantum potential. Journal of the Optical Society of America B, 40(3), 620-630.

\bibitem{Hojman and Asenjo (2020)} Hojman, S. A., \& Asenjo, F. A. (2020). A new approach to solve the one-dimensional Schrödinger equation using a wavefunction potential. Physics Letters A, 384(36), 126913.

\bibitem{Villaflor et al. (2024)} Villaflor, V. A., Muñoz-Mosqueira, V. A., \& Hojman, S. A. (2024). A new approach to solving the Schrödinger equation using wavefunction potentials in two and three dimensions. The European Physical Journal Plus, 139(5), 391.

\bibitem{Soto-Eguibar et al. (2021)} Soto-Eguibar, F., Asenjo, F. A., Hojman, S. A., \& Moya-Cessa, H. M. (2021). Bohm potential for the time dependent harmonic oscillator. Journal Of Mathematical Physics, 62(12).

\bibitem{Moya-Cessa et al. (2022)} Moya-Cessa, H. M., Hojman, S. A., Asenjo, F. A., \& Soto-Eguibar, F. (2022). Bohm approach to the Gouy phase shift. Optik, 252, 168468.

\bibitem{Louisell (1990)} Louisell, W. H. (1990). Quantum statistical properties of radiation. Wiley.

\bibitem{Loudon and Knight (1987)} Loudon, R., \& Knight, P. L. (1987). Squeezed light. Journal of modern optics, 34(6-7), 709-759.

\bibitem{Yuen (1976)} Yuen, H. P. (1976). Two-photon coherent states of the radiation field. Physical Review A, 13(6), 2226.

\bibitem{Caves (1981)} Caves, C. M. (1981). Quantum-mechanical noise in an interferometer. Physical Review D, 23(8), 1693.

\bibitem{Moya et al. (2014)} Moya-Cessa, H. M., Soto-Eguibar, F., \& Christodoulides, D. N. (2014). A squeeze-like operator approach to position-dependent mass in quantum mechanics. Journal of Mathematical Physics, 55(8).

\bibitem{Tapia-Valerdi et al. (2025a)} Tapia-Valerdi, M. A., Ramos-Prieto, I., Soto-Eguibar, F., \& Moya-Cessa, H. M. (2025). Waveguide arrays interaction to second neighbors: Exact solution. arXiv preprint arXiv:2501.12550.

\bibitem{Tapia-Valerdi et al. (2025)} Tapia-Valerdi, M. A., Ramos-Prieto, I., Soto-Eguibar, F., \& Moya-Cessa, H. M. (2025). Waveguide Arrays: Interaction to Many Neighbors. Dynamics, 5(3), 25.

\bibitem{Perez-Leija et al. (2010)} Perez-Leija, A., Moya-Cessa, H., Szameit, A., \& Christodoulides, D. N. (2010). Glauber–Fock photonic lattices. Optics letters, 35(14), 2409-2411.

\bibitem{Susskind and Glogower (1964)} Susskind, L., \& Glogower, J. (1964). Quantum mechanical phase and time operator. Physics Physique Fizika, 1(1), 49.

\bibitem{London (1926)} London, F. (1926). Über die Jacobischen transformationen der quantenmechanik. Zeitschrift für Physik, 37(12), 915-925.

\end{thebibliography}
\end{document}